\documentclass[aps,prb,twocolumn,citeautoscript,nofootinbib,eqsecnum,10pt,longbibliography]{revtex4-2}  
\usepackage{amsmath,amssymb,bm} 
\usepackage{graphicx}

\usepackage[tight]{subfigure} 
\usepackage{comment}
\usepackage{color} 
\usepackage[papersize={8.3in,11.7in}]{geometry}
\usepackage[colorlinks=true]{hyperref}
\hypersetup{
	bookmarks=true,         % show bookmarks bar?
	unicode=false,          % non-Latin characters 
	pdftoolbar=true,        % show Acrobat
	pdfmenubar=true,        % show Acrobat 
	pdffitwindow=false,     % window fit to page when opened
	pdfstartview={FitH},    % fits the width of the page to the window
	pdftitle={boundary corrections},    % title
	pdfauthor={Author},     % author
	pdfsubject={Subject},   % subject of the document
	pdfcreator={Creator},   % creator of the document
	pdfproducer={Producer}, % producer of the document
	pdfkeywords={keyword1} {key2} {key3}, % list of keywords
	pdfnewwindow=true,      % links in new window
	colorlinks=true,       % false: boxed links; true: colored links
	linkcolor=magenta, %red,          % color of internal links (change box color with linkbordercolor)
	citecolor=blue,        % color of links to bibliography
	filecolor=magenta,      % color of file links
	urlcolor=blue           % color of external links
} 

\usepackage{amsfonts,bbold}
\usepackage{upgreek}
\usepackage{slashed}
\usepackage{latexsym}
\usepackage{ulem}

\usepackage{wrapfig} % to include figure with text wrapping around it						
\usepackage{wasysym}			
\usepackage{amsthm}
\usepackage{xcolor}
\usepackage{nicefrac}

\newcommand{\nn}{\nonumber \\}
\newcommand{\be}{\begin{equation}}
\newcommand{\ee}{\end{equation}}

\renewcommand{\vec}[1]{{\bf #1}}

\def \bece {\begin{center}}
	\def \ence {\end{center}}
\def \bega {\begin{gathered}}
\def \enga {\end{gathered}}
\def \benu {\begin{enumerate}}
	\def \ennu {\end{enumerate}}
\def \beit {\begin{itemize}}
	\def \enit {\end{itemize}}
\def \bede {\begin{description}}
	\def \ende {\end{description}}
\def \betb {\begin{tabular}}
	\def \entb {\end{tabular}}
\def \bear {\begin{array}}
	\def \enar {\end{array}}

\makeatletter
\renewcommand*\env@cases[1][1.4]{%
	\let\@ifnextchar\new@ifnextchar
	\left\lbrace
	\def\arraystretch{#1}%
	\array{@{}l@{\quad}l@{}}%
}
\makeatother

\renewcommand{\epsilon}{\varepsilon}

\begin{document}

\title{Boundary logarithmic corrections to the dynamical correlation functions
of one-dimensional spin-$1/2$ chains}

\author{Imke Schneider$^1$}
\author{Ipsita Mandal$^2$}
\author{Polina Matveeva$^3$}
\author{Dominik Strassel$^1$}
\author{Sebastian Eggert$^1$}

\affiliation{$^1$Department of Physics and Research Center OPTIMAS,  University of 
	Kaiserslautern, 67663 Kaiserslautern, Germany\\
	$^2$Institute  of  Nuclear  Physics,  Polish  Academy  of  Sciences,  31-342  Krak\'{o}w,  Poland\\
	$^3$Department of Physics, Bar Ilan University, Ramat Gan 52900, Israel}

\begin{abstract}
The asymptotic dynamical correlation functions in one-dimensional spin chains are described by power-laws.  The corresponding exponents characterize different bulk and boundary critical behavior. We present novel results for the logarithmic contribution to the boundary correlations of an isotropic Heisenberg chain. The exponent of the logarithm, $\lambda=1$, is derived using a renormalization group technique. We confirm our analytical results by comparing with numerical quantum Monte Carlo data.
\end{abstract}

\date{\today}

\maketitle
%\tableofcontents

\section{Introduction}
\label{sec:intro}

The isotropic spin-$1/2$ chain is one of the most prominent 
examples of a quantum many-body system. It is fair to say that the
one-dimensional Heisenberg model has been an inspiration for fruitful 
theoretical developments for exact methods,
bosonization, and numerical algorithms, ever since the invention of the Bethe ansatz in the early days of quantum mechanics \cite{Bethe31}.
Critical exponents for spin-spin power-law correlations were first predicted by Luther and Peschel \cite{peschel} in 1975, which agreed with the pioneering numerical results of Bonner and Fisher \cite{bonner} from 1964. In 1989, it was realized that the marginally irrelevant ``spin-Umklapp'' operator leads to multiplicative corrections, with 
logarithmically {\it increasing} behavior in the asymptotic long-distance limit, and is of the form \cite{Affleck_1989,giamarchi1989,singh1989,affleck1998}
\begin{align}
\label{SzSz}
G_{zz}(x,y,t) & = \langle S^z(x,t) S^z(y,0) \rangle \nn
&= {\rm const.} \ 
%\frac{(-1)^x}{(2 \pi)^{3/2}} 
(-1)^{x-y}\, \frac{\sqrt{\ln r/r_0}}{r} \,,
\end{align}
where $r = \sqrt{(x-y)^2-v^2 t^2}$ is the space-time distance.  This was also confirmed
numerically in real space \cite{hallberg,eggert96}. In a similar calculation, the logarithmic corrections to the dimer correlations, due to the marginally irrelevant operator, were studied, 
and the asymptotic form of the correlation function was obtained \cite{Vekua2016, Hikihara2017}.

The focus of this paper is the corresponding logarithmic correction of the boundary
critical behavior.  Boundaries play an important role in one-dimensional systems. This is due to the fact that for an antiferromagnetic spin chain  impurities will
effectively cut the chain at low temperatures \cite{Kane92}, resulting in zero electric \cite{Kane92PRB} and magnetic conductance \cite{Eggert92}. Antiferromagnetic exchange anisotropies correspond to repulsive interactions in fermionic models.  Reflecting boundaries induce Friedel oscillations \cite{egger1995,eggert95,leclair1996,rommer2000, soeffing2009} and characteristic boundary correlations \cite{fabrizio1995} which have a large impact on the local density of states  in fermionic systems \cite{eggert1996,mattsson1997,schneider2008,schneider2010,soeffing2013} as well as the dynamical structure factor in doped spin chains \cite{bohrdt2018}.  Boundary thermodynamics for spin chains and the local susceptibility have been investigated earlier using field theory techniques \cite{susc,eggert2002,fujimoto2004,furusaki2004,Sirker2008,Sirker2009} and the quantum transfer matrix methods \cite{goehmann2005,kozlowski2012,pozsgay2018}. Here, we focus on the spin-spin correlation function for an isotropic chain, which has a different power-law 
behavior at the boundary \cite{Eggert92,affleck1999,brunel1999}, viz. 
\begin{equation}
G_{zz}(x=y=d,t) =  {\rm const.\ } 
%\frac{(-1)^x}{(2 \pi)^{3/2}} 
\frac{(\ln t/t_0)^\lambda}{t^2}\,,
\end{equation}
for spins close to a boundary (i.e., where $x=y$ 
are of the order of the lattice spacing $d$).
The exponent $\lambda$ of the logarithm was first predicted to be $\lambda=4$ in a preprint \cite{preprint}, but in subsequent works, $\lambda=2$ has been reported \cite{brunel1999,affleck1999}. Ref.~\cite{affleck1999} uses non-abelian bosonization, while the result of Ref.~\cite{brunel1999} has been derived implementing abelian bosonization.  In our paper, we argue that the exponent is $\lambda=1$ using abelian bosonization, and we also present numerical data based on quantum 
Monte Carlo simulations to support our analytical results.

%%%%%%%%%%%%%%%%%%%%%%%%%%%%%%%%%%%%%%
\section{Model and algebraic decay}

Before discussing the multiplicative logarithmic corrections, we first briefly summarize the form of the correlations described by the algebraically decaying power-laws, both in the bulk and near the boundaries.

The Hamiltonian of the anisotropic Heisenberg chain in terms of spin-$1/2$ operators  $\mathbf{S}_i=(S^x_i,S^y_i,S^z_i)$ at site $i$ reads
\begin{align} 
\label{model}
H=J \sum_{i=1}^{L-1} \left( S_i^xS_{i+1}^x +  S_i^y S_{i+1}^y  
+ \Delta  S_i^z S_{i+1}^z \right),
\end{align}
where $\Delta$ is the anisotropy parameter, and $L$ is the size of the system. Here, we consider  ``open''  boundary conditions, i.e., the edge spins at $i=1$ and $i= L$ are not coupled with each other. For $\Delta=1$, the Hamiltonian is invariant under SU(2) transformations, which is the parameter regime that we will consider in the subsequent discussions.  
For a more general description, we will use abelian bosonization \cite{Luttinger63,Haldane81,Giamarchi03}, as the 
mode expansion is known in this language, thus allowing an explicit calculation of the algebraically-decaying boundary correlation functions \cite{mattsson1997}. The SU(2) symmetry will be strictly maintained by enforcing the equality of the correlation functions for the transverse and longitudinal directions.

The low-energy effective bosonic description of the anisotropic Heisenberg chain is the Luttinger liquid Hamiltonian \cite{Luttinger63,Haldane81,Giamarchi03}
\begin{align}
\label{Hamiltonian0}
H  =  \frac{\upsilon}{2}\int_0^L dx & \left [   K \left(  \partial_x \tilde{\theta} \right)^2
+  \frac{1} {K} \left(  \partial_x \tilde{\phi} \right)^2  \right ],
\end{align}
where $\tilde{\phi}$ is the bosonic field, and $\partial_x\tilde{\theta}$ is its conjugate momentum, such that $[\tilde{\phi}(x),\partial_x\tilde{\theta}(y)]=i \,\delta(x-y)$.   The spin-wave velocity $v=J \,\pi \sin\vartheta/2 \vartheta$, and the Luttinger parameter $K=\pi/2(\pi-\vartheta)$ are known analytically as functions of $\cos\vartheta=\Delta$, from the exact solution
of the model \cite{Giamarchi03}. In our notation, $K=1/2$ corresponds to the SU(2)-invariant point $\Delta=1$. The Luttinger parameter $K$ controls the decay of the correlation functions. It can be gauged away by the canonical transformation
\begin{align}\label{can_trafo}
\phi=\tilde{ \phi}/\sqrt{K}, \quad \theta= \sqrt K\,\tilde{\theta}\,,
\end{align} 
which maps the above Hamiltonian onto a free boson Hamiltonian 
\begin{align}
H_0=\frac{v}{2}\int_0^L dx \left [ \left(  \partial_x \theta \right)^2 
+   \left(  \partial_x \phi \right)^2\right].
\end{align}
So far we have omitted the spin-Umklapp operator, which will be discussed in the following
section.

Algebraic correlation functions can be determined using the mode expansion (see also the Appendix \ref{obcbosonization}) \cite{Giamarchi03,eggert97}.  
The overall prefactor of the correlation functions depends on the choice of 
cutoff in the field theory, but for the 
spin model it can be fixed 
using exact methods \cite{Lukyanov03}.
For the field theory, it is useful to set the normalization such that in the 
thermodynamic limit $L\to \infty$, the two-point function of the vertex operator  
$ e^{ i \,\gamma \phi}$ (far from the boundary) has the form
\begin{align}
\label{cftnormalization}
\langle e^{i\gamma \phi(x,\tau)}
\, e^{- i  \gamma \phi(y,0)}\rangle
=\frac{1}{\left [\,r(x-y,\tau) \,\right ]^{2d_s}}\,,
\end{align}
for imaginary time $\tau=i \,t$. 
Here, $d_s=\frac{\gamma^2}{4 \pi}$ is the scaling dimension of the operator, and we introduced 
\begin{align}
r(x,\tau)&=\sqrt{x^2 + v^2 \,\tau^2 }
\end{align}
which denotes the  space-time distance.

In the following, we are interested in the dominant antiferromagnetic correlations, which 
arise from the alternating parts of the spin  operators in the bosonized form, given by
\begin{align}
\label{S_bosonised}
S^z(x,t)&= A\, (-1)^x \sin\left(\sqrt{4 \,\pi\, K} \,\phi(x,t)\right),\\
S^+(x,t)&=\tilde{A} \,(-1)^x \,e^{- i\,\sqrt{\frac{\pi}{K}}\,\theta(x,t)},
\label{Splus}
\end{align}
where $A$ and $\tilde{A}$ are related to the amplitudes 
of the asymptotic correlation functions \cite{Lukyanov03}. 
We are now in a position to calculate the 
longitudinal correlation function $G_{zz}(x,y,\tau)$
[cf. Eq.~(\ref{SzSz})], and the transverse correlation function
\begin{align}
G_{+-}(x,y,\tau)=\langle  S^+(x,\tau)  \,   S^-(y,0) \rangle /2 \,.
\end{align}
At the SU(2)-invariant point, the two correlation functions coincide, viz. $G_{zz}=G_{+-}$.

Since we are interested in correlations near boundaries, we
evaluate the expectation values using a finite-size bosonization approach,
where the mode expansions of the bosonic fields  are chosen such that the open boundary conditions 
of the system are fulfilled. More details are provided in the Appendices \ref{szszunperturbed} and \ref{spsmunperturbed}. Once the finite-size results are known, these can be generalized to a semi-infinite system with size $L\to\infty$, and with a boundary as $x\to 0$.
We find \cite{Eggert92,bohrdt2018} that
\begin{align}
& G_{zz}(x,y,\tau)
\nn &
= \frac{A^2 (-1)^{x+y} } { 2 \left ( 4\,x\,y  \right )^K }
\left [ \left \lbrace 
\frac{r(x+y,\tau)} {r(x-y,\tau)}\right \rbrace^{ 2K} 
%%%%
-  \left \lbrace \frac{r(x-y,\tau)} {r(x+y,\tau)}\right \rbrace^{2K} 
\right ] 
\label{szszalt}
\\
&  \simeq  \frac{A^2}{2} (-1)^{x+y} \times\begin{cases}
 r^{-2K} &\text{ bulk } \\ 
\frac{  \left( 4\, x\,y  \right)^{1-K} } { v^2 \tau^2} &\text{ boundary }
%\frac{  2 K x^2}{(x y)^{1+K}}  &\text{ (perp.~limit)}
\end{cases},
\label{szszbulkbdyperp}
\end{align}
for the Luttinger liquid Hamiltonian in Eq.~(\ref{Hamiltonian0}). The bulk limit refers to  $x\,y \gg (x-y)^2 + v^2\,\tau^2 $, while the boundary limit implies  $ x,\, y \ll v\, t $.  We have used $r\equiv r(x-y,\tau)$ without arguments to simplify notation.

An analogous calculation for the transverse correlation function yields
\begin{align}
& G_{+-}(x,y,\tau)
\nn &
= \frac{\tilde{A}^2}{2} (-1)^{x+y} \left(\frac{4\,x\,y}
%%%%%%
{  r^2(x+y,\tau) \,r^2(x-y,\tau)
}\right) ^  {\frac{ 1 } {4 K }}
\label{s+s-alt}
\\
&  \simeq 	\frac{\tilde{A}^2}{2} (-1)^{x+y}  \,
 \begin{cases}
r^{-\frac{ 1 }{2 K }}  &\text{ bulk} \\ 
\left(\frac{ 4\, x\,y }{ v^4\,\tau^4} \right)^{\frac{ 1 }
{4 K }} 
&\text{ boundary}
\end{cases}.
\label{spsmbulkbdyperp}
\end{align}

At the SU(2)-invariant point, the normalization factors $A$ and $\tilde A$ diverge \cite{Lukyanov03},
which is the first indication that the prefactors also become dependent on the distance.  Nonetheless, for later convenience, we ignore the overall normalization 
and introduce the short notation $G^0(x,y,\tau)$ for the power-law 
correlation functions at the isotropic point with $K=1/2$. 
The corresponding asymptotic power-law decays in the two limits follow directly from  Eqs.~(\ref{szszbulkbdyperp}) and (\ref{spsmbulkbdyperp}) as
\begin{align}
\label{G0}
G^0(x,y,\tau)&\propto
 (-1)^{x+y} \times\begin{cases}
r^{-1} &\text{ bulk} \\ 
\frac{2 \sqrt{ x\,y}   } {v^2\,\tau^2}  &\text{ boundary}
%\frac{  \sqrt{x}}{y^{\frac{3}{2}}}  &\text{ (perp.~limit)}
\end{cases}.
\end{align}

%%%%%%%%%%%%%%%%%%%%%%%%%%%%%%
\section{Renormalization group flow}

Multiplicative logarithmic corrections from the spin-Umklapp operator 
were first derived using non-abelian bosonization \cite{Affleck_1989} and 
also using abelian bosonization \cite{giamarchi1989, singh1989}, as two independent approaches. The second approach is the one we will use here for the boundary case.
The spin-Umklapp operator 
is of the form $\cos\left(\sqrt{16 \,\pi\, K} \,\phi\right)$.  Its scaling dimension  $d_s=4\,K$  changes continuously with $K$. In particular, at $K=1/2$, the operator is marginal -- hence the corrections
from a renormalization group (RG) approach are only logarithmically small
at best, and must therefore be treated with great care. For this purpose, we will expand the Hamiltonian in Eq.~(\ref{Hamiltonian0}) around the
SU(2)-invariant point with $K=1/2$.
Using the notation $K=1/2 +\delta K$ and subsequently rescaling the fields according to Eq.~(\ref{can_trafo}) we get
\begin{align} 
H=H_0
+   \int_0^L dx \left [\, g_1\, \mathcal{O}_1 (x)
+g_2 \,\mathcal{O}_2(x)\, \right ]
\label{Hamiltonian1}
\end{align}
with
\begin{align}
\mathcal{O}_1 & = \frac{v}{4}
\left[ \left(  \partial_x \theta \right)^2 
-  \left(  \partial_x \phi \right)^2\right ], 
\,\, \mathcal{O}_2= \frac{v} {2\,\pi}\cos\left(  \sqrt{ 8 \,\pi}\,\phi \right),
\label{eqcorr}
\end{align}
and $g_1=4\,\delta K$, while $g_2$ is the coupling constant of the spin-Umklapp operator. The value of $g_2$ relies on the chosen normalization of the bosonic vertex operators, which in our case is the field theory normalization in Eq.~(\ref{cftnormalization}).
 %%%%%%%%%%%%%%%%%%%%%%%%
This Hamiltonian defines the starting point of our analysis.
The aim of this section is to explain the RG technique to treat the perturbing $\cos \left( \sqrt{8\,\pi\,} {\phi} \right)$ operator in abelian bosonization. We will first revisit the mechanism how the logarithmic corrections to the correlation functions at the SU(2)-invariant point are derived in the bulk limit. We will then employ this analysis to a system with open boundary conditions. 

The behavior of the bulk theory  under renormalization is
 well-known. The RG flow equations, which describe how the couplings $g_1$ and $g_2$ evolve under a change of the length scale $\Lambda$, are of the Kosterlitz-Thouless type \cite{Kosterlitz_1974,affleck_leshouches1988,giamarchi1989}. The dependence of the coupling constants on the relevant length scale is encoded in their derivatives with respect to the logarithm of the scale, which, for historical reasons are called beta-functions. In our notation, these read
 \begin{align}
 \label{betafunctions}
 \frac{dg_{2}}{dl} =- { g_1\,g_2 } \,,
 \quad 
 \frac{dg_{1}}{dl} =- g_{2}^2 \,,
 \end{align}
 where $l=\ln (\Lambda/\alpha_0) $. The parameter $\Lambda$ denotes 
the physical length scale at which the system is studied, i.e., the corresponding energy scale serves as the infrared cutoff. The ultraviolet energy cutoff corresponds to the length scale $\alpha_0$, which has been estimated to be slightly smaller than the lattice spacing $d$, i.e., $\alpha_0\approx 0.85\,d$ \cite{eggert96}. 
The SU(2)-invariant point corresponds to
 \begin{align}
 g_1=g_2=g   \,,
 \end{align} 
 where the two beta-functions coincide. Eq.~(\ref{betafunctions}) can be exactly solved, and for the isotropic point, the solution is given by
 \begin{align}
 g(l) &=\frac{g_0} { 1+b \,g_0 \,\ln \left(\Lambda/\alpha_0\right ) }\,.
 \label{solution_betafunc}
 \end{align}
Here, $b=1$ and $g_0$ is the bare coupling, when $\Lambda$ is of the order of $\alpha_0$.

In order to derive the logarithmic corrections to the correlation functions, we will follow the approach of Refs.~\cite{giamarchi1989,singh1989}. 
The multiplicative corrections to the unperturbed correlation function are captured by a function $F_{\mu\nu}(x,y,\tau)$, which is defined by the equation
\begin{align} 
\label{Gab}
G_{\mu\nu}(x,y, \tau)=G^0(x,y,\tau)\, F_{\mu\nu}(x,y,\tau)\,,
\end{align}
where the subscript $\mu \nu$ is the label for the ``$zz$'' or the ``$+-$'' components of the correlations.
Along the entire line $g_1=g_2=g$ isotropy must hold [thus implying $F_{zz}(x,y,\tau)
=F_{+-}(x,y,\tau) $], and for
vanishing $g_1=g_2=0$ the multiplicative function must be equivalent to the identity [thus implying $F_{\mu\nu}(x,y,\tau)=1$]. 
The factor $F_{\mu\nu}$ can be derived from the leading-order  corrections of the perturbation theory. Here we use the interaction representation \cite{Abrikosov65} and the imaginary time  $\tau=i \,t$. The first-order perturbative correction in $g_1$ and $g_2$ then reads \cite{Abrikosov65} 
\begin{align}
G_{\mu\nu}(x,y,\tau) =  & G^0(x,y,\tau) + 
 \sum_{i=1}^2 {\mathcal{T}}_{\mathcal{O}_i}^{\mu\nu}(x,y,\tau) \,,
\label{eqrgcorr0}
\end{align}
where 
\begin{align} 
&{\mathcal{T}}_{\mathcal{O}_i}^{\mu\nu}(x,y,\tau) \nn
& = - g_i f_{\mu \nu}\int_0^\infty d\tilde x \int_{-\infty}^\infty  d\tilde \tau  \Big \langle {T} \,
S^\mu(x,\tau)S^\nu(y,0)
%& \hspace{3.5 cm}
\mathcal{O}_i(\tilde x , \tilde  \tau)  \Big  \rangle\,\nonumber\\ 
& + g_i\, G^0(x,y,\tau)\int_0^\infty d\tilde x \int_{-\infty}^\infty d\tilde \tau  \Big \langle
\mathcal{O}_i(\tilde x , \tilde  \tau)  \Big  \rangle .
\label{Tab}
\end{align}
with $f_{zz}=1$ and $f_{+-}=\frac{1}{2}$. 
The symbol ${T}$ in the expectation value denotes the time-ordering operator. 
In the above expression, the time integral is part of the interaction representation, while the interaction Hamiltonian itself is an integral over the space variable (where we have set the upper boundary $L\to\infty$, valid for a semi-infinite system).
The second term in Eq.~(\ref{Tab}) represents the disconnected diagrams, which we subtract off in order to cancel the unphysical singular contributions.  
In the next section, we will discuss the expectation values of ${\mathcal{T}}_{\mathcal{O}_1}^{\mu\nu}$  and ${\mathcal{T}}_{\mathcal{O}_2}^{\mu\nu}$ for open boundary conditions, and evaluate the integrals under a change of cutoff.
Combined with the knowledge of the relation between the bare and renormalized couplings, this will determine the factor $F_{\mu\nu}$. 

We will illustrate the renormalization of the correlation functions, and how a multiplicative factor $F_{\mu\nu}$ emerges considering the bulk case. Here, $F_{\mu\nu}=F_{\mu\nu}(r/\alpha_0)$ [cf. Eq.~(\ref{Gab})] is a function of the ratio $r/\alpha_0$ only.
The explicit calculations of ${\mathcal{T}}_{\mathcal{O}_1}^{\mu\nu}$  
and ${\mathcal{T}}_{\mathcal{O}_2}^{\mu\nu}$ (demonstrated in the next section) will show 
that the integrals, which determine the perturbed correlation function in Eq.~(\ref{Tab}), exhibit diverging parts around singular points. The latter need to be regularized by the cutoff $\alpha_0$.  
The integrands take the general form 
\begin{align}
f(\tilde{x},\tilde{\tau},x,y,\tau)
\Bigg / \left[\prod_s r^2(\tilde{\vec r}-\vec r_s) \right ],
\end{align}
 where $\tilde{\vec r}=(\tilde{x},\tilde{\tau})$ is the  variable of integration, $\vec r_s=(x_s,\tau_s) $ denotes a singular point, and $f(\tilde{x},\tilde{\tau},x,y,\tau)$ is a polynomial function [cf.  Eq.~(\ref{SzSzapp}), shown in the later part of the paper]. In the bulk limit, the integral structure further simplifies to involve only two singularities at $\vec r_1=(x,\tau)$ and $\vec r_2 = (y,0)$. The elementary integral that needs to be solved is given by \cite{singh1989} 
 \begin{align}
 I=\iint^\prime d \tilde{x} \,d \tilde{\tau} 
 \frac{1}{r^2(\tilde{\vec r}-\vec r_1)\, r^2(\tilde{\vec r}-\vec r_2)} \,.
 \end{align}
We regularize the two-dimensional integral by excluding
 circles of radii $\alpha_0$ around $\vec r_1$ and $\vec r_2$ while performing the integrations, which is indicated by the symbol $\iint^\prime$.  While carrying out the integrations, we
choose the spatial coordinate axis to be along $\vec r_1-\vec r_2$. In the polar coordinates, the integral can then be evaluated
 \begin{align}
 	I &=\int_{}^{'}  
d\tilde{r} \int_0^{2\pi} d\theta \,\frac{1}{\tilde{r} 
\left (\tilde{r}^2-2\, \tilde{r} \,r \cos \theta + r^2 \right )} \nn
 &= \frac{4\,\pi}{r^2} \ln\left(\frac{\Lambda}{\alpha_0}\right),
 	\label{elementaryInt}
 \end{align}
where we have subtracted the singularities within radii $\alpha_0\ll r$ 
around $\vec r_1$ and $\vec r_2$, after doing the angular integration exactly.
Here, $\Lambda \approx r/\sqrt{2}$ is given in terms of
the distance 
$r= r(x-y,\tau)$  
between the two singular points. Note that the distance $r$ serves as the infrared cutoff $\Lambda$ of the integral, because the integrand decays as $1/\tilde{r}^3$ for $\tilde{r} \gg r$. 
In the following, the upper cutoff of the RG procedure is therefore always limited by $r$, which determines the end-point of the logarithmic RG behavior. Hence, each singularity contributes a term $2 \,\pi \ln\left(\frac{\Lambda}{\alpha_0}\right)$ times the remaining integrand.
This logarithm  gives a small correction to the correlation function, as long as $\Lambda/\alpha_0\sim 1$. In this case, the perturbed correlation function can be generally written as
\begin{align}\label{form_firstorder}
G_{\mu\nu}=G^0 \left[1+ \left (a_1\, g_1+a_2 \,g_2  \right ) 
\ln\left(\frac{\Lambda}{\alpha_0}\right)\right],
\end{align}
where $g_1=g_2=g(l)$, and $a_1$ and $a_2$ are constants resulting from the perturbative contributions of $\mathcal{O}_1$ and $\mathcal{O}_2$, respectively.   
With increasing $\Lambda\sim r$, the ``corrections'' become arbitrarily large, and consequently, the perturbative approach seems to be doomed.  On the other hand, 
Eq.~(\ref{form_firstorder}) remains correct if we only want to consider a small
change in the cutoff $\Lambda\to \Lambda'$.
Since the underlying field theory is scale-invariant, Eq.~(\ref{form_firstorder}) can always be 
used to calculate the perturbative correction corresponding to a small change $dl = \ln\left(\frac{\Lambda'}{\Lambda}\right)$ in the values of the cutoff. Of course, the coupling constant $g(l)$ is {\it not} scale-invariant, as it depends on the overall value $l=\ln\left(\frac{\Lambda}{\alpha_0}\right)$ in Eq.~(\ref{solution_betafunc}), which must be taken into account at each RG step.
We therefore take the cutoff $\Lambda$ as a tunable variable, 
which can be increased step-by-step from an 
initially small value $\Lambda=\alpha_0$, until the physical cutoff $\Lambda\sim r$ is reached. At each step, we use Eq.~(\ref{form_firstorder}) to calculate 
the correction of $G_{\mu\nu}$, assuming that 
$l=\ln\left(\frac{\Lambda}{\alpha_0}\right)$ only changes 
by an infinitesimal amount $dl = \ln\left(\frac{\Lambda'}{\Lambda}\right)$, and 
$ g_1=g_2=g(l) $ is given by the running coupling constants in Eq.~(\ref{solution_betafunc}).
To make it more concrete, let us look at the renormalization 
of $F_{\mu\nu}$ as we slightly increase the cutoff $\Lambda\to \Lambda'$ . This is captured by
\begin{align}\label{Fabdef}
F_{\mu\nu}(\Lambda'/\alpha_0) = 
F_{\mu\nu}(\Lambda/\alpha_0)
\left[1+a \,g(l) \ln\left(\frac{\Lambda'}{\Lambda}\right)\right],
\end{align}
where $a = a_1 + a_2$.  
Iterating and multiplying all RG steps from  $\Lambda=\alpha_0$ to $\Lambda=r$, we find
that 
\begin{align}
\label{Fab}
F_{\mu\nu}(r/\alpha_0)
&=\prod_{l=0}^{ \ln (r/\alpha_0)} \left[1+a \,g(l)dl \right] \nonumber\\
&= \exp\left[\int_0^{\ln (r/\alpha_0)} \gamma_{\mu\nu}(l)\,dl\right],
\end{align}
where we have introduced
\begin{align} 
\label{gamma}
\gamma_{\mu\nu}(l)= a\, g(l)\,,
\end{align} 
which is twice the commonly defined anomalous dimension of the corresponding spin-field \cite{Amit78}. Notice that the anomalous dimension is defined as the logarithmic derivative of the prefactor, which renormalizes the correlation function multiplicatively under a change of scale \cite{Amit78}. Therefore, it depends on the coupling of the theory at any scale.    
Finally, making use of Eq.~(\ref{solution_betafunc}), and performing the integral in Eq.~(\ref{Fab}), we obtain the form of multiplicative logarithmic correction as
\begin{align}
F_{\mu\nu}(r/\alpha_0)
&=\exp\left\{   \frac{a}{b} \ln\left[1+b \,g_0 \ln (r/\alpha_0)  \right]   \right\}\nn
&= \left[1+b \,g_0 \ln (r/\alpha_0)\right]^{\frac{a}{b}}\,.
\label{logcorrgeneral}
\end{align}
Eq.~(\ref{logcorrgeneral}) thus provides a general recipe for the logarithmic correction to any correlation function. It involves two characteristic quantities: (1) the beta-function 
of the theory which determines the parameter $b$; and (2) the anomalous dimension $\gamma_{\mu\nu}$
of the corresponding spin-field in Eq.~(\ref{gamma}), which determines the parameter $a$. 

Indeed, the concept of the anomalous dimension is well known from the
Callan-Symanzik equation, describing the evolution of any $n$-point correlation function under variation of the energy scale \cite{ZinnJustin}, which of course gives an identical 
result \cite{Affleck_1989,giamarchi1989,singh1989}.
In the case of open boundary conditions, however, the above step-by-step RG treatment 
appears to be more transparent, because  there are two length scales in the boundary theory,   
viz.~$r(x-y,\tau)$ and $r(x+y,\tau)$. 
We will show that in the bulk and boundary limits, the two length scales reduce again
to a single one, viz.~$\Lambda=r$ and $\Lambda=v \,t$, respectively. In these cases, the RG treatment is fully analogous to the one described above, with the corresponding length scales taken into account.
We will now proceed to explicitly determine the parameters $a$ and $b$ entering Eq.~(\ref{logcorrgeneral}), for both the bulk and the boundary limits. 
%using a first order perturbation theory. 

%%%%%%%%%%%%%%%%%%%%%%%%%%%%%%%%%%%%%%%%%%%%%%%
\section{First order perturbation}

In this section, we discuss the first order corrections to the free correlation function, due to the presence of the operators $\mathcal{O}_1$ and $\mathcal{O}_2$. We identify the logarithmically divergent pieces of the resulting  integrals, in order to obtain the general form  of Eq.~(\ref{form_firstorder}), and to determine the prefactors $a_1$ and $a_2$.  The knowledge of these prefactors is crucial as they enter the final exponent of the multiplicative logarithmic exponent in Eq.~(\ref{logcorrgeneral}) where $a=a_1+a_2$.  Our  focus will be on the boundary behavior of a semi-infinite chain. For a better understanding, we will also include results for the bulk limit, where we recover the known results for the infinite case.

Let us start by discussing the contribution of the operator  $\mathcal{O}_1$. In this case, we do not need to evaluate the expectation value in a first order perturbative expression at all, as there is a much simpler way to determine the constant $a_1$. The operator $\mathcal{O}_1$ can be included in the quadratic part of the Hamiltonian, allowing it to be treated exactly. This just affects the value of the Luttinger parameter, which increases as $K\to K+\delta K$ with $\delta K=g_1/4$. Eq.~(\ref{szszbulkbdyperp}) for the correlation function $G_{zz}(x,y,\tau)$  is still  valid, using 
$K=\frac{1}{2}+\delta K$ in the vicinity of the isotropic point. The first order correction   $ {\mathcal{T}}_{\mathcal{O}_1}^{zz}$ can now be obtained by expanding the power-law expression for $G_{zz}(x,y,\tau)$ in $\delta K$, leading to
\begin{equation}
\label{T1zzbulkbdy}
{\mathcal{T}}_{\mathcal{O}_1}^{zz}
%\nn 
=-  \frac{  g_1\,A^2\, (-1)^{x+y} }  {2}  
\begin{cases}
\frac{1}{2\,r} \ln \left( \frac{\Lambda}{ \alpha_0}\right)&\text{ bulk}\\
0 & \text{ boundary}\\
\end{cases}.
\end{equation}
Here, we have used the fact that the infrared cutoff is given by the distance $\Lambda\sim r$, 
relative to the ultraviolet cutoff $\alpha_0$. Note that in the boundary case, the power $(x\, y)^{1-K}$ appearing in the correlation function in Eq.~(\ref{szszbulkbdyperp}), contributes a logarithm  $\sim\ln 
\left( {x \,y}\,/{\alpha_0^2} \right)$, which vanishes near the boundary 
$x, y\sim\alpha_0$ -- hence it has been neglected here.

For the contribution of the operator  $\mathcal{O}_2$, we need to evaluate the integral corresponding to the first order perturbation explicitly, which is a straightforward but cumbersome calculation. 
We can express the perturbative change 
of the correlation function 
from ${\mathcal{O}_2}$
as a sum of the two integrals $I_1$ and $I_2$, such that
\begin{align}
\label{pertT2}
& {\mathcal{T}}_{\mathcal{O}_2}^{zz}(x,y,\tau)  \\ &
= - \frac{g_2 \,v \, A^2 (-1)^{x+y}} {8\,\pi\, \sqrt{x \, y}} 
\left[\frac{ r(x+y,\tau)   } {r(x-y,\tau)}\, {I}_1
	- \frac{r(x-y, \tau)}  { r(x+y,\tau) }  \, {I}_2 \right] .\nonumber
	\end{align}
Here,
\begin{equation}
{I}_i=\iint^\prime  d \tilde x \, d \tilde \tau 
\, t_i(x,y,\tilde x,\tau,\tilde{\tau})\,,
\label{intI}
\end{equation}
for $i=1,2$, and the integral therefore denotes the correction due to a 
small change of the cutoff around each singularity.
Each integrand $t_i$ is determined from evaluating the time-ordered 
correlation function in Eq.~(\ref{Tab}), using the mode expansion shown in the Appendix \ref{obcbosonization}. This gives us (for details see Appendix \ref{perturbationzz})
\begin{align}
\label{SzSzapp}
&  t_{1/2} = \\
&\left[ \frac{  (\tilde{\tau}^2 + \tilde{x}^2+y^2) x \mp ((\tau - \tilde{\tau})^2  + \tilde{x}^2 +x^2) y }
%&\left[ \frac{  r^2(\tilde x+y,\tilde\tau) \,x \mp r^2(\tilde x+x,\tilde \tau-\tau) y }
%&\left[ \frac{  r^2(\tilde{x}+y,\tilde{\tau}) \,x \mp r^2(\tilde{x} \pm x,\tilde{\tau}-\tau)  \,y }
{r({\tilde x} - x, \tilde \tau-\tau )\, r({\tilde x} 
+ x, \tilde\tau- \tau ) \,r({\tilde x} - y, \tilde \tau )
\,r({\tilde x} + y, \tilde \tau )}  
\right]^2  . \nonumber 
\end{align}
Since the correlations are symmetric under $\tilde x \to -\tilde x$, we do not need
to restrict the integration to positive values only.
In the above expression, the contributions from the disconnected diagrams have already been subtracted. The final integrals are dominated 
by the behavior of the integrands in the vicinity of the singular points,  
which  determine the leading order logarithmic contributions. 
For both the terms $t_1$ and $t_2$, there are four singular points, which are located at  $(\tilde x \rightarrow \pm x , \,  \tilde \tau \rightarrow \tau )$, and $(\tilde x \rightarrow \pm y , \, \tilde \tau \rightarrow 0)$.

 We now determine the leading logarithmic contributions to the integrals, by choosing an appropriate parametrization in the vicinity of each singular point. For instance, we choose $(\tilde x= x +\delta{\tilde x}, \, \tilde \tau = \tau+\delta \tilde \tau)$ for the point  $(\tilde x \rightarrow x, \, \tilde \tau \rightarrow \tau )$, and then expand for small $\delta\tilde x$ and $\delta \tilde \tau$. Analogous to Eq.~(\ref{elementaryInt}), each singularity contributes
$2 \, \pi \ln(\Lambda/\alpha_0)$ times the corresponding value of the remaining integrand, leading to 
 \begin{align}
   I_1 &\simeq 2\,\pi 
  \left[ \frac{r(x-y,\tau)} {r(x+y,\tau)}\right]^2  
  \ln\left(\frac{\Lambda}{\alpha_0}\right )  , \\
   I_2 &\simeq   2\,\pi 
 \left[ \frac{r(x+y,\tau)} {r(x-y,\tau)}\right]^2   
  \ln\left(\frac{\Lambda}{\alpha_0}\right)  .
  \end{align}
Inserting these results in Eq.~(\ref{pertT2}), we obtain
	\begin{align}
\label{T2zzfull}
& {\mathcal{T}}_{\mathcal{O}_2}^{zz} \\
& = \frac{  g_2 \, A^2 \,(-1)^{x+y}} {4 \,\sqrt{x \,y} } 
\left[
  \frac{ r(x+y,\tau) } { r(x-y,\tau) } 
-\frac{ r(x-y,\tau) } { r(x+y,\tau) } 
\right]
\ln \left( \frac{\Lambda} {\alpha_0}\right) \nonumber
\end{align}
which reduces  to
\begin{align}
\label{T2zzbulkbdy}
& {\mathcal{T}}_{\mathcal{O}_2}^{zz} \\
&= \frac{ g_2\, A^2 \,(-1)^{x+y}} {2} 
\begin{cases}
\frac{1}{r} \ln \left( \frac{\Lambda}{ \alpha_0}\right)&\text{ bulk}\\
\frac{2\, \sqrt{x\,y}   } { v^2\,\tau^2}
\ln \left( \frac{\Lambda} { \alpha_0} \right)& \text{ boundary}
\end{cases},\nonumber 
\end{align}
in the two limits. 
Finally, we add  ${\mathcal{T}}_{\mathcal{O}_1}^{zz}$ in Eq.~(\ref{T1zzbulkbdy}), and ${\mathcal{T}}_{\mathcal{O}_2}^{zz}$ in Eq.~(\ref{T2zzbulkbdy}), to the  unperturbed correlation function in  Eq.~(\ref{G0}). This leads to
\begin{align}
\label{firstordercorr}
& G_{zz}(x,y,\tau) \\
&=\frac{A^2\,(-1)^{x+y}}  {2}  
\begin{cases}  
\frac{1}{ r}
\left[1+\left(g_2
-\frac{  g_1} { 2}\right)
\ln \left( \frac{\Lambda}  { \alpha_0 }\right)\right] & \text{ bulk} \\
\frac{2 \sqrt{x\, y}} {v^2 \,\tau^2}
\left [ 1+ g_2 
\ln \left( \frac{\Lambda}  { \alpha_0 }\right)\right] 
&\text{ boundary}
\end{cases}.\nonumber
\end{align}
We thus obtain an expression which agrees with the general form shown in  Eq.~(\ref{form_firstorder}), with the coefficients ($a_1=-\frac{1}{2}, a_2=1$, $a=\frac{1}{2} $) in the bulk limit, and ($a_1=0$, $a_2=1$, $a=1 $) in the boundary limit. 

We are now in the position to state the final result for the exponent $\lambda=\frac{a}{b}$ 
of the logarithmic correction, both in the bulk and boundary limits. 
Using Eq.~(\ref{firstordercorr}) and Eq.~(\ref{logcorrgeneral}),
we find that $a=\frac{1}{2}$ and $ b=1$ in the bulk limit. This gives us the logarithmically corrected correlation function \cite{Affleck_1989,giamarchi1989, singh1989}
\begin{equation}
G(r)=\frac{A^2 (-1)^{x+y}} {2\, r} \sqrt{\ln \left(  r/\alpha_0\right )  }\,,
\end{equation} 
with a bulk logarithmic exponent $\lambda= {1}/{2}$.
%%%%%%%%%%%%%%%%%%%%%%%%%%%%%%%%%
In the boundary limit, we find that $a=1$ and $ b=1$, which 
results in the logarithmically corrected correlation function
\begin{equation}
G_{b}(t) =-\frac{A^2 (-1)^{x+y}\,\sqrt{x\, y}}{v^2 \, t^2 }  
\ln \left( \frac{ t} {   t_0 }\right) \,
\end{equation} 
with a boundary logarithmic exponent $\lambda=1$.
This is the main result of this paper, which, however,
does not agree with what was obtained earlier \cite{affleck1999,brunel1999}. Hence, it warrants a critical discussion about the possible origin of this discrepancy. 

%%%%%%%%%%%%%%%%%%%%%%%%%%%%%%%%
\section{Critical discussion}

In this section, we will explain the reasons why we believe that we do not recover the boundary logarithm exponent $\lambda=2$  of previous works \cite{affleck1999,brunel1999}. While we have obtained the same beta-functions, and the coefficient $b=1$, we obtain a different anomalous dimension with the coefficient $a=1$ (as opposed to $a=2$, found in earlier works). The  discrepancy stems from taking a different order of the calculational steps -- in our case, we first evaluate 
the full integral in Eq.~(\ref{T2zzfull}),  and then 
take the boundary limit in Eq.~(\ref{T2zzbulkbdy}).
In contrast, in the earlier papers \cite{affleck1999,brunel1999}, where an
operator product expansion was employed, the order of calculations 
amounts to first taking the boundary limit in the correlation function, for small $x,y$. However, when taking the boundary limit, the singularities (appearing at $\tilde{x}=-x$ and $\tilde{x}=x$) get partially reduced due to factors in the numerator, leaving only one singularity instead of two (albeit with a prefactor which is four times larger).
Note that this procedure assumes that $x,y \alt \alpha_0$, and therefore 
does not capture the diverging 
dependence on the lower length scale cutoff $\alpha_0$ correctly. On the other hand, in our approach, the arguments $(x,y)$ are always larger than the cutoff $\alpha_0$, as they must be, judging from physical intuition and the estimate $\alpha_0\approx 0.85\,d$ \cite{eggert96}.  In the bulk limit, in contrast, the order of the computational steps does not make any difference. In fact, the bulk limit can be safely taken before integrating, as all singularities are always sufficiently removed from each other. 

Let us show explicitly that the expansion for small $(x,y)$ as the first step, starting from the 
{\it same} correlation function in Eq.~(\ref{SzSzapp}), leads to a different result.
First expanding ${\mathcal{T}}_{\mathcal{O}_2}^{zz}$ in Eq.~(\ref{pertT2}) for small $(x,y)$, before performing the integral, we find that
\begin{align}
& {\mathcal{T}}_{\mathcal{O}_2}^{zz}(x,y,\tau) \\
=&\frac{g_2 \,v \, A^2 (-1)^{x+y}} {2\,\pi} 
\iint^\prime  d \tilde x \, d \tilde \tau  \frac{ \sqrt{x\,y}}{r^2(\tilde{x},\tilde{\tau}) \,r^2(\tilde{x},\tilde{\tau}-\tau)}\,. \nonumber
\end{align}
Here, we have dropped all terms which are odd under parity [viz., $\tilde x \to -\tilde x$], and which therefore do not contribute to the integral. 
The integrand is singular for $(\tilde x \to 0, \tilde \tau \to 0 )$ and $(\tilde x \to 0, \tilde \tau \to \tau)$. Integration in the same manner as before yields
\begin{align}
{\mathcal{T}}_{\mathcal{O}_2}^{zz}(x,y,\tau)
=g_2  \, A^2 (-1)^{x+y} \; \frac{ 2 \sqrt{x\,y}}
{ v^2 \, \tau^2} \ln\left(\frac{\Lambda}{\alpha_0}\right).
\end{align}
Note the additional factor 2 as compared to Eq.~(\ref{T2zzbulkbdy}). This factor leads to an anomalous dimension $a=2$ and as a result the logarithmic exponent $\lambda$ is twice as large as our exponent $\lambda=1$ for the boundary case. Since the analytic structure is changed by taking the limit of small $(x,y)$ first, we do not believe that the previous result of $a=2$ is correct.

A further check of our result comes from the transverse correlation function, which has an identical logarithmic correction. A calculation analogous to the one before for the contribution of the $\mathcal{O}_1$ operator, expanding $K=\frac{1}{2}+\delta K$ for small $\delta K$ in the transverse correlation function, gives us
\begin{equation}
{\mathcal{T}}_{\mathcal{O}_1}^{+-}
=\frac{ g_1\, \tilde{A}^2}{2} \,(-1)^{x+y}\times
\begin{cases}
\frac{1}{2r} \ln \left( \frac{\Lambda}{ \alpha_0}\right)&\text{ bulk}\\
\frac{ 2\sqrt{x\,y} }{ v^2\,\tau^2} \ln \left( \frac{\Lambda}{\alpha_0}\right) & \text{ boundary} 
\end{cases}.
\end{equation}
For this case, the operator $\mathcal{O}_2$ does not generate a logarithmically divergent term in first order in $g_2$ as shown in the Appendix \ref{perturbationpm}. In other words, 
\begin{align}
{\mathcal{T}}_{\mathcal{O}_2}^{+-}=0\,.
\end{align}
This calculation confirms once again that $a=1$ in the boundary case, as also required 
by the rotational invariance. Even though the separate contributions of $a_1$ and $a_2$ are different for the transverse and longitudinal correlations functions, the sum $a=a_1+a_2$ is the same. Indeed, we find that 
along the isotropic line $g_1=g_2=g$, the logarithmic corrections to 
the decay of the spin-spin correlation functions follow
\begin{align}
\label{firstordercorrspsm}
& G_{+-}(x,y,\tau) 
= \frac{\tilde{A}^2}{2} \,(-1)^{x+y} \\
& \,\times \begin{cases}  \frac{1}{r}
\left[1+
\frac{ g_1} { 2}
\ln \left( \frac{\Lambda}  { \alpha_0 }\right)\right] & \text{bulk} \\
\frac{2 \sqrt{x \,y}} {v^2 \,\tau^2}
\left[ 1+ g_1 
\ln \left( \frac{\Lambda}  { \alpha_0 }\right)\right] &\text{boundary}
\end{cases}, \nonumber
\end{align}
and thus
\begin{align}
{\mathcal{T}}_{\mathcal{O}_1}^{zz}+{\mathcal{T}}_{\mathcal{O}_2}^{zz}={\mathcal{T}}_{\mathcal{O}_1}^{+-}+{\mathcal{T}}_{\mathcal{O}_2}^{+-}.
\end{align}

%%%%%%%%%%%%%%%%%%%%%%%%%%%%%%%%
\section{Numerical data and fits}

%%%%%%%%%%%%%%%%%%%%%%%%
\begin{figure}[t]
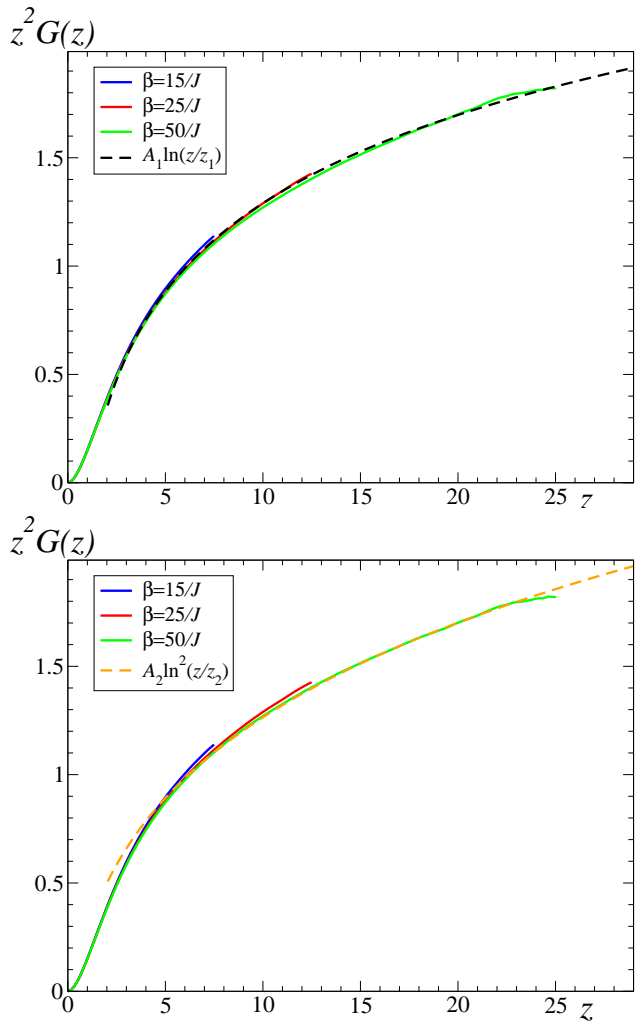

%\subfigure[]{\includegraphics[width = 0.6 \columnwidth]{fit.eps}}\hspace{0.25 cm}
%\subfigure[]{\includegraphics[width = 1. \columnwidth]{fit1.eps}}\hspace{0.25 cm}
%\subfigure[]{\includegraphics[width = 1.  \columnwidth]{fit2.eps}}
{\includegraphics[width = 1. \columnwidth]{fit1.eps}}\\
{\includegraphics[width = 1.  \columnwidth]{fit2.eps}}
\caption{The curves show the data points obtained from the quantum Monte Carlo Simulations for the boundary correlation function $z^2 \,G_{zz}(z)$ of the last spin in an isotropic chain in imaginary-time as a function of $z= \frac{v \,\beta}{\pi} 
\sin \left( \frac{\pi \tau}{\beta} \right)$. The three curves correspond to  $\beta J =15, 25, 50$.  The fits according to Eq.~(\ref{fit}) with the parameter values taken from the last line of the Table~\ref{tab} are shown for $\lambda=1$ (top) and $\lambda=2$ (bottom).
\label{data}}
\end{figure}
%%%%%%%%%%%%%%%%%%%%%%%%%%%%

In order to test the predictions form our analytical calculations, we have performed numerical quantum Monte Carlo simulations, using the Stochastic Series Expansion algorithm \cite{Sandvik92,Sengupta02} with directed loop updates 
\cite{Sandvik02b} and a Mersenne Twister random number generator \cite{Mersenne}. 
With this method, it is straight-forward to calculate correlation functions in the
imaginary time $\tau$ at finite temperatures \cite{Sandvik92}.

Using the mode expansion of the fields, the method to calculate the power-law correlations 
for any finite size $L$ and finite temperature $T$ is well-known \cite{mattsson1997,eggert97,eggert2002}.
In particular, for finite temperatures, the imaginary space-time 
coordinate $i \,x+ v \,\tau$ in
the correlation functions is replaced by \cite{susc}
\begin{equation}
i \,x+ v \,\tau \to z=\frac{v\, \beta}{\pi} 
\sin \Big(\frac{\pi \left (i \,x + v\,\tau \right )}
{v \, \beta} \Big)\,.
\end{equation}
Here, it is assumed that the inverse temperature is $\beta=1/T \ll L/v$, such that
the effects from the finite system size $L$ can be ignored \cite{mattsson1997,eggert97,eggert2002}.
It has been argued in Ref.~[\onlinecite{Barzykin_1999}] that the RG treatment can also be
performed in the variable $z$, up to second order in the beta-function, 
but that higher order RG may not give a perfect data collapse as a function of $z$.
We, therefore, analyze the imaginary-time correlation function of the last 
spin of an open chain, using the ansatz
\begin{equation}
G_{zz}(x=y=d,\tau) =  {\rm A_\lambda\ } 
\frac{[ \,\ln \left(z/z_\lambda\right)\, ]^\lambda}{z^2} \,,
\label{fit}
\end{equation}
with $z= \frac{v \,\beta}{\pi} \sin \left(\frac{\pi\, \tau}{\beta} \right)$,
exponent $\lambda= 1, 2$, and 
fitting constants $A_{1,2}$ and $z_{1,2}$.
The corresponding quantum Monte Carlo data for an isotropic chain of length $L=500$, at three different 
temperatures $\beta J =15, 25, 50$
are shown in Fig.~\ref{data}, where we have multiplied the correlation function with the 
leading power-law $z^2$. The results for different temperatures have approximately 
the same functional dependence on $z$, and hence the higher order RG corrections,
which are not functions of $z$, can indeed be neglected.

\begin{table}  
	\begin{tabular}{|l|l|l|l|l|}
\hline	
$\beta J$ & $A_1$ & $z_1$ & $A_2$ & $z_2$\\
\hline	
	15 & 0.583235& 0.935679&0.104749 & 3.65832\\ \hline
	25 & 0.584947 &0.908307 &0.0864403 &4.72989 \\ \hline
	50 & 0.594767 &0.86585 &0.070273 &6.78464 \\ \hline
	all &0.58763 &0.898285 &0.0674096 &7.60085 
\\\hline	
	\end{tabular} 
	\caption{Fitting parameters.}\label{tab}
\end{table}
%%%%%%%%%%%%%%%

The constants $A_{1,2}$ and $z_{1,2}$ have been determined by the best fits to Eq.~(\ref{fit})
for $z > 2$ and are shown in Table \ref{tab}. At the first sight, both $\lambda=1$ and $\lambda=2$
appear to fit well with the data. But upon closer inspection, the fit for $\lambda =2$ has
clear systematic deviations, as seen in Fig.~\ref{data} and Table~\ref{tab}. In particular, the fitting parameters in Table~\ref{tab} drift as a function of $\beta$ by 30\% or more, while the parameters for $\lambda=1$
remain constant for different $\beta$ within a few percent. Notice that it is important to analyze the quality of the fit over the entire range of data for different temperatures.
We therefore conclude that the data shows clear 
evidence  that $\lambda=1$ is the correct exponent for the boundary logarithmic corrections.

%%%%%%%%%%%%%%%%%%%%%%%%%%%%%%%
\section{Conclusion}
 We have considered the boundary logarithmic correction to the dynamical spin-spin correlation function, for the isotropic Heisenberg chain. The logarithmic corrections are caused by the marginal spin-Umklapp operator, which we have treated by RG techniques. When both spins are located close to the open boundary, the correlations are captured by a single scale $\upsilon t$, which allows using scaling arguments analogous to the bulk theory. We have found an anomalous dimension which results in an exponent $\lambda=1$ for the logarithmic correction. This result is confirmed by the state-of-the-art numerical data from quantum Monte Carlo simulations. 
 
The time-dependent spin-spin correlation functions can be probed by the nuclear magnetic resonance (NMR) relaxation rates of spin-$1/2$ antiferromagnetic chain compounds. For the bulk materials, experimental data on Sr$_2$CuO$_3$ \cite{Takigawa1996} have been found to be in good agreement with theoretical predictions incorporating multiplicative logarithmic corrections \cite{Takigawa1997,Barzykin2001}. In addition, impurity effects on the NMR spectra have also been studied \cite{Takigawa1997imp}. Recently, the magnetic properties of doped spin chains have attracted renewed experimental interest \cite{Utz2017, Karmakar2017}. As the current experimental resolution for NMR data is much better than the numerical capabilities to simulate the behaviour at long times, we believe that a comparison of our results to the boundary NMR relaxation rates  \cite{brunel1999} will potentially resolve the issue. An alternative (and extremely promising) route is given by experiments on ultracold gases realizing the spin-1/2 Heisenberg chain \cite{Fukuhara2013, Jepsen2020}, where measurements of time and
space-resolved correlations have been proposed \cite{bohrdt2018,Knap2013,bohrdt2018ultracold}.  As these systems always have finite system sizes, precise knowledge of the boundary effects, that we have investigated in our paper, is of great importance.

\begin{acknowledgments}
We are grateful for useful discussions with J. Sirker and F. G\"ohmann.   This work was supported by the Deutsche Forschungsgemeinschaft (DFG) via the SFB/Transregio 185, projects A4 and A5, and the Forschungsgruppe FOR 2316 project P10. IM acknowledges the warm hospitality of TU Kaiserslautern during the calculations for this paper.
\end{acknowledgments}

%%%%%%%%%%%%%%%%%%%%%%%%

\begin{appendix}

\begin{widetext}	

\section{Mode Expansions for the semi-infinite chain}	
\label{obcbosonization}

To account for the open boundary condition on the left end of the semi-infinite chain, we impose the Dirichlet boundary condition on the spin field at the boundary $x=0$ \cite{brunel1999, Eggert92}, which implies that $ \left \langle S^z(x=0) \right \rangle =0$. By using the expression for the bosonized spin operators
[cf. Eq.~(\ref{S_bosonised})], where  the bosonization field $\phi(x,t) =   \phi_L (x,t) + \phi_R (x,t)$, we obtain $\phi_0 \equiv \phi (x=0) 
=k \,\sqrt{\pi/\left( 4K\right)}$ (with $k \in \mathbb{Z} $). Since the chiral bosonic fields $\phi_R$ and $\phi_L$ are functions of $(x-v\, t)$ and $(x+v \,t)$, respectively, the Dirichlet boundary condition allows us to relate them as $\phi_R (x,t) = -\phi_L ( -x,t) + \phi_0$.
Hence, the mode-expansions for these fields take the form \cite{Eggert92,eggert97}:
\begin{align}
	\label{formulas_OBC}
	& \phi_L (x,t) =\frac{ \phi_0+\tilde{\phi}_0    } {2} +\frac{x+v\,t}{2\,L}\,Q +  
\sum_{\ell > 0}\left[ \frac{  i\,e^{ -\frac{i\,\pi \,\ell} {L} (x + v\,t)}}
		{ \sqrt{ 4\, \pi\,\ell} } \, b_{\ell}  
	+  \mathrm{H.c.} \right]\,, \nn 
	%%%%%%%%%%%%
	&\phi_R (x,t) 
		=\frac{ \phi_0 - \tilde{\phi}_0    } {2} +\frac{x -v\,t}{2\,L}\,Q  
+    \sum_{\ell > 0}\left[- \frac{  i\,e^{ \frac{i\,\pi \,\ell} {L} (x - v\,t)}}
	{ \sqrt{ 4\, \pi\,\ell} }  \,b_{\ell}  +  \mathrm{H.c.} \right]\,, \nn
	%\sum_{n > 0} \frac{e^{- \frac{\pi\, i \,n} {L} (x-v\,t)}}
%	{ \sqrt{ 4\, \pi\,n} }  b_{n}^\dagger  \,,\nn
	%%%%%%%%%%%%%%%%%%%%%%
	& \phi(x,t)  =  {\phi}_0  +\frac{ x }{ L}\,Q
	  + \sum_{\ell > 0} \frac{\sin \left (\frac{\pi\,\ell \,x} {L} \right )}
	{ \sqrt{\pi\,\ell} } \left[ e^{ -\frac{i\,\pi \,\ell\,v\,t} {L}  }\, b_{\ell} + \mathrm{H.c.} \right]\,, \nn
%\mathrm{	i}\, e^{ \frac{\pi\, i \,n\,v\,t} {L}  } \, b_{n}^\dagger  \right ),\nn
	%%%%%%%%%%%%%%%%%%%%%%%%%%%%%5
	& [b_\ell, \,b_{\ell^\prime}^\dagger]=\delta_{\ell,\ell^\prime}\,, \quad [\tilde \phi_0, \,Q] = i\,.
	\end{align}

For the conjugate field $  \theta(x,t) =  \phi_L (x,t) - \phi_R (x,t) $ we have
\begin{align}
	\theta(x,t)  = \tilde{\phi}_0  +\frac{ v\,t}{ L}\,Q
	+    \sum_{\ell > 0} \frac{\cos \left (\frac{\pi\,\ell \,x} {L} \right )}
	{ \sqrt{\pi\,\ell} } \left[i\, e^{ -\frac{i\,\pi\,\ell\,v\,t} {L}  }\, b_{\ell}+ 
	\mathrm{H.c.} \right].
	\label{thetamode}
	\end{align}

\section{Unperturbed correlation function $G_{zz}(x,y,t)$  }
\label{szszunperturbed}

 In this section, we will explain how to calculate expectation values in general, using the mode expansion of the previous section. An example of this procedure is given by the evaluation of the unperturbed correlation function $G_{zz}(x,y,t)$. 
 
 We start from the bosonized spin operator 
 \begin{align}
 S^z(x,t)=A\,(-1)^x \sin\left(\sqrt{4\pi K}\,\phi(x,t)\right) ,
 \end{align}
as shown in Eq.~(\ref{S_bosonised}). The expectation values in the ground state can be evaluated in the simplest way by normal ordering all the expressions, with respect to the annihilation and creation operators $b_\ell$ and $b_\ell^\dagger$, respectively.  Using the Baker-Campbell-Hausdorff formula  $e^{A+B} =e^A\,e^B 
\,e^{-\frac{1}{2}[A,B]}$ for $\left [A,[A,B]\right]=0$ and $\left [B,[B,A] \right ]=0$, we find that, for a general vertex operator \cite{bohrdt2018},
\begin{align}
  e^{i\,\alpha\,\sqrt{4\pi}\,\phi(x,t)}
& = \left[ \frac{   \pi  }
{ 2\,L \sin\left (   \pi\,x /L   \right )}
\right ]^{\alpha^2} 
\exp\left(i\, \alpha \sum_\ell e^{i\,\omega_\ell \,t} \,\frac{ A_\ell^\dagger(x)}
{\sqrt{\ell}}\right) \exp\left(i \,\alpha \sum_\ell e^{-i\,\omega_\ell\, t} 
\,\frac{ A_\ell(x)}{\sqrt{\ell}} \right) \\
& = \left[ \frac{   \pi  }
{ 2\,L \sin\left (   \pi\,x /L   \right )}
\right ]^{\alpha^2} 
:e^{i\,\alpha\,\sqrt{4\pi}\,\phi(x,t)}:\,,
\label{corr01}
\end{align}
where $\omega_\ell=\frac{\pi v \,\ell}{L}$ and
\begin{align}
A_\ell(x)=2  \sin\left(\frac{\ell \,\pi \,x }{L}\right) b_\ell \,.
\end{align}
Here, we have used the expansion of the logarithm $\sum_{\ell=1 }^\infty \frac{e^{- i \omega_\ell t}}{\ell}
=-\ln \left(1-e^{-\frac{ i \pi  v t}{L} }\right)$ where we assume that $t$ comes with a small negative imaginary part to ensure convergence. We also have set the normalization of the single vertex operator such that for the two-point function far away from the boundary Eq.~(\ref{cftnormalization}) is fulfilled. 
For a product of two vertex operators, we further need to normal order the inner products of the exponentials, in order to obtain a fully normal ordered expression. This yields
\begin{align}
   e^{i\,\alpha\,\sqrt{4 \pi}\phi(x,t)}\, e^{ i\,\beta\,\sqrt{4 \pi}\phi(y,0)}  
=  \frac{\left(\frac{   \pi }{ 2\,L }\right)^{\alpha^2+\beta^2}
  e^{  - \alpha\,\beta \,C_1 (x,y,t)}  }{\sin^{\alpha^2}\left (   \frac{ \pi\,x } {L }  \right )\sin^{\beta^2}\left (   \frac{ \pi\,y } {L }  \right ) } : e^{i\,\alpha\,\sqrt{4 \pi}\phi(x,t)}\, e^{ i\,\beta\,\sqrt{4 \pi}\phi(y,0)}  :\,,
\label{corr02}
\end{align}
with the commutator 
\begin{align}
 C_1(x,y,t) & =  \sum_{\ell,\ell^\prime}
 \frac{e^{-i \, \omega_\ell \, t}}
 {\sqrt{\ell \,\ell^\prime}}[A_\ell(x),A^\dagger_{\ell^\prime}(y)]=4\sum_\ell \frac{e^{-i \, \omega_\ell \, t}}{ \ell} \sin \left(\frac{\ell\,\pi x }{L}\right)\sin \left(\frac{\ell\,\pi y }{L}\right) \\
& =
\Big [
-\ln \left( 1- e^{- \frac{    i\pi}  {L}(x-y+v\,t) }  \right)
+ \ln\left( 1- e^{- \frac{  i\pi}  {L}(x+y+v\,t) }   \right)
\nn & \qquad \quad 
+ \ln\left( 1- e^{ \frac{    i\pi}  {L}(x+y - v\,t) }   \right)
- \ln \left( 1- e^{ \frac{    i\pi}  {L}(x-y- v\,t) }  \right)\Big]
\nn
%%%%%%%%%%%%%
& \simeq
\ln\left[
\frac{  \left( x+y \right)^ 2-(v \,t )^2
 }
 %%%%%
{  \left( x-y \right)^ 2-(v \,t)^2
 }
\right ] .
\end{align}
 In the last line, we have taken the limit $L\to \infty$.
Thus, we get
\begin{align}
e^{i\,\alpha\,\sqrt{4 \pi}\, \phi(x,t)}\, e^{ i\,\beta\,\sqrt{4 \pi}\, \phi(y,0)}  
 \simeq \frac{\left[
	\frac{  \left( x+y \right)^ 2-v^2 \,t^2}
	%%%%%
	{  \left( x-y \right)^ 2-v^2 \,t^2}
	\right ]^{-\alpha \,\beta} }  {(2 \,x)^{\alpha^2} (2 \,y)^{\beta^2}  } : 
	e^{i\,\alpha\,\sqrt{4 \pi}\, \phi(x,t)}\, e^{ i\,\beta\,\sqrt{4 \pi}\, \phi(y,0)}  : \,.
\end{align}
For the correlation function 
\begin{align}
G_{zz}(x,y,t)=A^2 \,(-1)^{x+y}
\left \langle  \sin\left(\sqrt{4\pi K}\, \phi(x,t)\right)\sin \left(\sqrt{4\pi K}\, \phi(y,0)  
\right) \right \rangle,
\end{align}
we need to combine four different products of the vertex operators. Using the above expressions, to leading order,
we find that
\begin{align}
& G_{zz}(x,y,t)
=
 \frac{A^2  \,(-1)^{x+y}} 
 { 2 \left (4 \,x \,y  \right )^K }
\left [  \left \lbrace \frac{  \left( x+y \right)^ 2-v^2 \,t^2}
{  \left( x - y \right)^ 2-v^2 \,t^2} \right \rbrace ^{K}-\left \lbrace 
\frac{  \left( x-y \right)^ 2-v^2 \,t^2}
{  \left( x+y \right)^ 2-v^2 \,t^2}\right \rbrace ^{K} 
%%%%
\right ] ,
\end{align}
which agrees with Eq.~(\ref{szszalt}) after switching to imaginary time (i.e., $t \to -i\, \tau$). 
In the bulk limit of $x\,y \gg  r^2$ [where $r^2 = (x-y)^2 - v^2\,t^2 $], we obtain
\begin{align}
%\label{szszbulk}
 G_{zz}(x,y,t) \big \vert_{\text{bulk}} & = \frac{A^2 \,(-1)^{x+y}} 
 { 2 \left (4 \,x \,y  \right )^K }
\left [  
 \left \lbrace \frac{   r^2 + 4\,x\,y }
{   r^2  } \right \rbrace ^{K} - \left \lbrace\frac{  r^2}
{   r^2 + 4\,x\,y }\right \rbrace ^{K} \right ] \simeq \frac{A^2 (-1)^{x+y}} 
{ 2 \left (4\, x\,y  \right )^K }
\left[ \left ( \frac{   4\, x\,y }
{   r^2  } \right )^{K} -0\right] \nn
%%%%%%%%%%%%%%%%
& 
 =
 A^2 \,(-1)^{x+y}\frac{1}{2\,r^{2K}} \,.
\end{align}
Finally, the boundary limit implies  $ x,\, y \ll v\, t $, which gives
\begin{align}
 G_{zz}(x,y,t) \big \vert_{\text{boundary}} & = - A^2\, (-1)^{x+y} 
 \frac{ \left (4 \,x \,y \right )^{1-K}} {2 \,v^2 \,t^2} \,.
 \end{align}

%%%%%%%%%%%%%%%%%%%%%%%%%%%%%%%%%%%%
\section{Unperturbed correlation function $G_{+-}(x,y,t )$}
\label{spsmunperturbed}

 In this section, we consider the spin operator
\begin{align}
S^+(x,t)=\tilde{A} \,(-1)^x\, e^{-i\,\sqrt{\frac{\pi}{K}}\, \theta(x,t)} \,,
\end{align}
as shown in Eq.~(\ref{Splus}), and calculate the unperturbed correlation function
 \begin{align}
 G_{+-}(x,y,t)=\frac{\tilde{A}^2\, (-1)^{x+y}}{2}
 \left \langle e^{-i\,\sqrt{\frac{\pi}{K}} \,\theta(x,t)} 
 \,e^{i\, \sqrt{\frac{\pi}{K}} \, \theta(y,0)} \right \rangle .
 \end{align} 
In this case, the relevant vertex operator in normal ordered form reads
\begin{align}
  e^{i\,\alpha\,\sqrt{4\pi}\, \theta(x,t)}
& = \left[ \frac{  2 \pi \,  \sin\left (   \pi\,x /L   \right )}
{ \,L }\right]^{\alpha^2} 
\exp\left(i \,\alpha \sum_\ell e^{i\,\omega_\ell\, t}
\, \frac{ B_\ell^\dagger(x)}
{\sqrt{\ell}}\right) \exp\left(i\, \alpha \sum_\ell e^{-i\,\omega_\ell \,t} \,
\frac{ B_\ell(x)}
{\sqrt{\ell}} \right),
%\label{corr01}
\end{align}
where 
\begin{align}
B_\ell(x)=2\,i \cos\left(\frac{\ell\,\pi \, x}  {L}\right) b_\ell\enspace.
\end{align}
For a product of two such operators, we obtain
\begin{align}
   e^{i\,\alpha\,\sqrt{4 \pi}\,\theta(x,t)}\, 
  e^{- i\,\alpha\,\sqrt{4 \pi}\, \theta(y,0)}  
= \left(\frac{  2 \pi   } { \,L }\right)^{2 \,\alpha^2}  
\sin^{\alpha^2}\left (   \frac{ \pi\,x } {L }  \right )\,
\sin^{\alpha^2}
\left (   \frac{ \pi\,y } {L }  \right ) \,
  e^{  \alpha^2 \,C_2 (x,y,t)}  : e^{i\,\alpha\,\sqrt{4 \pi}\, \theta(x,t)}\, 
 e^{- i\,\alpha\,\sqrt{4 \pi}\,\theta(y,0)}  :\,,
%\label{corr02}
\end{align}
with the commutator 
\begin{align}
 C_2(x,y,t) & = \sum_{\ell,\ell^\prime}
 \frac{e^{-i\, \omega_\ell \,t}}
 {\sqrt{\ell \, \ell^\prime}}[B_\ell(x),B^\dagger_{\ell^\prime}(y)]
 = 4\sum_\ell \frac{e^{-i \, \omega_\ell \, t}}{ \ell} 
 \cos \left(\frac{\ell\,\pi x}{L}\right)\cos \left(\frac{\ell\,\pi y}{L}\right) \\
& =
-\Big [
\ln \left( 1- e^{- \frac{  i \pi}  {L}(x-y+v\,t) }   \right)
+ \ln\left( 1- e^{- \frac{  i \pi}  {L}(x+y+v\,t) }   \right)
\nn & \qquad \quad 
+ \ln\left( 1- e^{ \frac{    i \pi}  {L}(x+y - v\,t) }   \right)
+ \ln \left( 1- e^{ \frac{    i \pi}  {L}(x-y- v\,t) }   \right)\Big]
\nn
%%%%%%%%%%%%%
& \simeq
-\ln\left[\frac{ \pi^4
 \left \lbrace \left( x+y \right)^ 2-(v \,t  )^2\right \rbrace
 %%%%%
 \left\lbrace 
  \left( x-y \right)^ 2-(v \,t )^2 \right \rbrace}
{L^4}  \right].
\end{align}
Note that the mode expansion of the $\theta$-field in Eq.~(\ref{thetamode}) includes the operator-valued zero mode $\tilde{ \phi}_0$. For a nonzero expectation value, such zero modes must cancel, which will restrict the possible combinations of the vertex operators. A further phase shift, resulting from the commutator of $\tilde{ \phi}_0 $ and $Q$, vanishes in the thermodynamic limit, and will be neglected here. 
Hence, for $L\to \infty$, we get
\begin{align}
e^{i\,\alpha\,\sqrt{4 \pi}\, \theta(x,t)}\, e^{ -i\,\alpha\,\sqrt{4 \pi}\, \theta(y,0)}  
&=   \left ( 4 \,x \,y\right )^{\alpha^2} \left[ 
 \left \lbrace \left( x+y \right)^ 2-v^2 \,t^2\right \rbrace 
%%%%% 
 \left \lbrace 
  \left( x-y \right)^ 2-v^2 \,t^2 \right \rbrace \right]^{-\alpha^2} 
% \nonumber \\  & 
: e^{i\,\alpha\,\sqrt{4 \pi}\,\theta(x,t)}\, 
e^{ -i\,\alpha\,\sqrt{4 \pi}\,\theta(y,0)}  :\,,
\end{align}
and setting $\alpha=-\sqrt{\frac{1}{4 K}}$, we obtain
\begin{align}
G_{+-}(x,y,t)=\frac{\tilde{A}^2\, (-1)^{x+y}} {2}
\left[ \frac{4\, x\, y}  {\left \lbrace \left (x+y\right )^2
-v^2 \, t^2 \right \rbrace 
\left \lbrace 
\left (x-y \right )^2-v^2 \,t^2 \right \rbrace }\right]^{\frac{1}{4K}}
\end{align}
as stated in the main text [cf. Eq.~$(\ref{s+s-alt})$], after employing $t\to -i\, \tau$.

% % % % % % % % % % % % % % % % % % % % % % % % % % % % % % % % % % % % % % % % % % %
\section{One-loop correction for $G_{zz}(x,y,\tau)$ from $\mathcal{O}_2$ }\label{perturbationzz}

In this section, we consider the one-loop correction for $G_{zz}(x,y,\tau)$, obtained from $\mathcal{O}_2(x,\tau)=\frac{v}{2\pi}\cos \big(\sqrt{8\pi}\,\phi(x,\tau) \big)$ and ${\mathcal{T}}_{\mathcal{O}_2}^{z z}$,
as shown in Eqs.~\eqref{eqrgcorr0} and \eqref{Tab}.  
The connected part of the time-ordered correlation function is  
\begin{align}
& \frac{2\pi}{v}  \left \langle T \,
S^z(x,\tau)\,S^z(y,0)\,
\mathcal{O}_2 (\tilde x , \tilde  \tau)  \right  \rangle_{\rm con} 
\nn & =     A^2 \,(-1)^{x+y}  \left \langle
\sin \left (\sqrt{2\pi}\, \phi (x,\tau)  \right ) 
\sin\left   (\sqrt{2\pi}\, \phi (y,0)  \right ) 
 \cos  \left ( \sqrt{8\pi}\, \phi(\tilde{x}, \tilde{\tau})  \right ) \right \rangle
-  G^{0}(x,y,\tau)  \left \langle 
\cos\left(\sqrt{8\pi}\,\phi (\tilde{x},\tilde{\tau})\right)\right \rangle ,
\label{corrcon}
\end{align} 
which we will evaluate here in real time.

As a first step, let us simplify the generic operator
\begin{align}
  e^{ i\,\alpha\,\sqrt{4\pi}\,\phi(x,t)} \, e^{ i\,\beta\,\sqrt{4\pi}\,\phi(y,0)}  
  \, e ^{i\, \gamma\,\sqrt{4\pi}\,\phi (\tilde{x},\tilde{t}) } 
 & \simeq 
 \frac{1 }
 { (2\,x)^{\alpha^2} (2 \,y)^{\beta^2} (2 \,\tilde{x})^{\gamma^2}
  }  
 \left[
\frac{  \left( x+y \right)^ 2-v^2 \,t^2
}
%%%%%
{  \left( x-y \right)^ 2-v^2 \,t^2
}
\right ]^{-\alpha \beta}
\left[
\frac{  \left( \tilde{x}+y \right)^ 2-v^2 \,\tilde{t}^2
}
%%%%%
{  \left(\tilde{ x}-y \right)^ 2-v^2 \,\tilde{t}^2
}
\right ]^{-\beta \,\gamma} \nonumber \\
%%%
 %%%%
& \quad \times \left[
\frac{  \left( \tilde{x}+x\right)^ 2-v^2 \,(\tilde{t}-t)^2
}
%%%%%
{  \left( \tilde{x}-x \right)^ 2-v^2 \,(\tilde{t}-t)^2
}
\right ]^{-\alpha\, \gamma}: e^{ i\,\alpha\,\sqrt{4\pi}\,\phi(x,t)} 
\, e^{ i\,\beta\,\sqrt{4\pi}\, \phi(y,0)}  \, 
e ^{i\, \gamma\,\sqrt{4\pi}\, \phi (\tilde{x},\tilde{t}) } :\,.
\label{threevertex}
%%%%%%%%%%%%%%%%%%%%%%%%%%%%%%%%%%%%%%%%%%%%%%%%%%%%%%%%%%%%%%%%%%%%%%%%%%%%% 
\end{align}
%%%%%%%%%%%%%%%%%%%%%%%%%%%%%%
Writing the $\sin(\sqrt{2\pi}\phi)$ and $\cos (\sqrt{2\pi}\phi)$ operators in the Euler forms, and using Eq.~(\ref{threevertex}) by setting $\alpha =\pm  1/{\sqrt{2}}$, $\beta =\pm 1/{\sqrt{2}}$, we find
that the  correlation function is given by
\begin{align}
& A^2 \,(-1)^{x+y}  
\left \langle  \sin \left (\sqrt{2\pi}\, \phi (x,t)  \right ) 
 \sin\left   (\sqrt{2\pi}\, \phi (y,0)  \right )  
 \cos  \left ( \sqrt{8\pi}\, \phi(\tilde{x}, \tilde{t})  \right ) \right \rangle   \nn 
%%%%%%%%%%%%%%%%%%%%%%%%%%%%%%%%%
 & = 
 \frac{  A^2 \,(-1)^{x+y}}
 { 32\, \sqrt{x \,y} \,\tilde{x} ^2  }  \,\Bigg \lbrace  
 %%%%%%%%%%%%%%%%%%%%%%%%%%%%%%%%%%%%%%%%%%%%
  \sqrt{
\frac
{  \left( x+y \right)^ 2-v^2 \,t^2
 }
{  \left( x-y \right)^ 2-v^2 \,t^2
 }
 }\cdot
%%%
\frac{  \left( \tilde{x} +y \right)^ 2-v^2 \,\tilde{t}^2
 }
 %%%%%
{  \left( \tilde{x}-y \right)^ 2-v^2 \,\tilde{t}^2
 }\cdot 
 %%%
 \frac{  \left( \tilde{x} -x \right)^ 2-v^2 \left ( \tilde{t} -t \right )^2
 }
{  \left( \tilde{x}+x \right)^ 2-v^2  \left ( \tilde{t} -t \right )^2
 }
 %%%%%%%%%%%%%%%%%%%%%%%
  \nn & \hspace{2.75 cm}
 -\sqrt{ \frac
 {  \left( x-y \right)^ 2-v^2 \,t^2
  }
 {  \left( x+y \right)^ 2-v^2 \,t^2
  }
  }
 %%%
 \cdot
 \frac
 {  \left( \tilde{x}-y \right)^ 2-v^2 \,\tilde{t}^2
  }
  {  \left( \tilde{x} +y \right)^ 2-v^2 \,\tilde{t}^2
  }\cdot
  %%%
  \frac
  {  \left( \tilde{x} -x \right)^ 2-v^2 \left ( \tilde{t} -t \right )^2
  }
 {  \left( \tilde{x}+x \right)^ 2-v^2  \left ( \tilde{t} -t \right )^2
  } 
    \Bigg \rbrace  +  (\tilde{x}\to -\tilde{x})\,.
\end{align}
For the  disconnected part, we have 
\begin{align}
   G^{0}(x,y,t) \left \langle \cos\left(\sqrt{8\pi}\,\phi (\tilde{x},\tilde{t})\right)\right \rangle 
&= \frac{\,A^2 (-1)^{x+y}} { 16 \,\sqrt{ x\, y} \,\tilde{x}^2}
\left \lbrace  \sqrt{\frac{  \left( x+y \right)^ 2-v^2 \,t^2}
{  \left( x - y \right)^ 2-v^2 \,t^2} }-\sqrt{ 
\frac{  \left( x-y \right)^ 2-v^2 \,t^2}
{  \left( x+y \right)^ 2-v^2 \,t^2}}
%%%%
\,\right \rbrace  .
\end{align}

Combining the above two terms, we get
\begin{align}
& \frac{2\pi}{v}\Big \langle T \,
S^z(x,\tau) \,S^z(y,0)
\mathcal{O}_2(\tilde x , \tilde  \tau)  \Big  \rangle_{\rm con} \nn
& = \frac{  A^2 (-1)^{x+y}}  { 32\,\, \sqrt{x \,y} \,\tilde{x} ^2  }  
\,\Bigg [ 
 %%%%%%%%%%%%%%%%%%%%%%%%%%%%%%%%%%%%%%%%%%%%
  \sqrt{\frac{  \left( x+y \right)^ 2-v^2 \,t^2}
{  \left( x-y \right)^ 2-v^2 \,t^2
 } }
%%%
\Bigg \lbrace \frac{  \left( \tilde{x} +y \right)^ 2-v^2 \,\tilde{t}^2
 }
 %%%%%
{  \left( \tilde{x}-y \right)^ 2-v^2 \,\tilde{t}^2}\cdot 
 %%%
 \frac{  \left( \tilde{x} -x \right)^ 2-v^2 \left ( \tilde{t} -t \right )^2
 }
 %%%%%
{  \left( \tilde{x}+x \right)^ 2-v^2  \left ( \tilde{t} -t \right )^2
 }-1 \Bigg \rbrace \nn
%%%%%%%%%%%%%%%%%%%% 
 & \hspace{ 2.5 cm}  -\sqrt{\frac
 {  \left( x-y \right)^ 2-v^2 \,t^2  }
 {  \left( x+y \right)^ 2-v^2 \,t^2 }}
 %%%
 \Bigg \lbrace
 \frac
 {  \left( \tilde{x}-y \right)^ 2-v^2 \,\tilde{t}^2 }
  {  \left( \tilde{x} +y \right)^ 2-v^2 \,\tilde{t}^2 }\cdot
  %%%
  \frac
  {  \left( \tilde{x} -x \right)^ 2-v^2 \left ( \tilde{t} -t \right )^2 }
 {  \left( \tilde{x}+x \right)^ 2-v^2  \left ( \tilde{t} -t \right )^2 }
 -1 \Bigg \rbrace  \Bigg ]  +(\tilde{x}\to -\tilde{x}) \,.
\end{align}
This expression can be rewritten as
\begin{align}
& \frac{2\pi}{v}\Big \langle T\,
S^z(x,\tau) \, S^z(y,0)
\mathcal{O}_2(\tilde x , \tilde  \tau)  \Big  \rangle_{\rm con} 
= \frac{  A^2 (-1)^{x+y}}{ 2\, \sqrt{x\, y}   }  \,\Bigg \lbrace  
%%%%%%%%%%%%%%%%%%%%%%%%%%%%%%%%%%%%%%%%%%%%
\sqrt{
	\frac
	{  \left( x+y \right)^ 2-v^2 \,t^2
	}
	{  \left( x-y \right)^ 2-v^2 \,t^2
	}
}
%%%
%%%
\,t_1 
-\sqrt{
	\frac{  \left( x-y \right)^ 2-v^2 \,t^2}
	{  \left( x+y \right)^ 2-v^2 \,t^2}
}
%%%
%%%
\,t_2  \Bigg \rbrace \,,
\end{align}
with 
\begin{align}
t_{1/2}
= \frac{ \left[
\left(-v^2\, \tilde{t}^2+\tilde{x}^2+y^2\right)x\mp 
\left \lbrace -v^2\, (\tilde{t}-t)^2
+\tilde{x}^2+x^2\right  \rbrace y \right]^2}
{\left\lbrace (\tilde{x}-y)^2-v^2\, \tilde{t}^2\right \rbrace
 \left\lbrace (\tilde{x}+y)^2-v^2 \,\tilde{t}^2\right \rbrace
\left \lbrace (\tilde{x}-x)^2 -v^2\, (\tilde{t}-t)^2\right \rbrace 
\left \lbrace (\tilde{x}+x)^2-v^2\,(\tilde{t}-t)^2\right \rbrace}\,.
\end{align}
This agrees with Eq.~(\ref{pertT2}) in the main text, after implementing $t \to -i\, \tau$. Note that the additional $1/2$-factor in Eq.~(\ref{pertT2}) results from extending the area of integration to the full plane.

 \section{One-loop correction for  $\langle  S^+(x,t)  \,   S^-(y,0) \rangle$ from $\mathcal{O}_2$}
 \label{perturbationpm}

 The one-loop correction to $\frac{1}{2}\langle  S^+(x,t)  \,   S^-(y,0) \rangle$ from $\mathcal{O}_2$, which determines ${\mathcal{T}}_{\mathcal{O}_2}^{+-}$ in Eq.~(\ref{Tab}), is given by
 \begin{align}
 \frac{\pi}{v}\Big \langle T \,
 S^+(x,\tau) \,S^-(y,0)\,
 \mathcal{O}_2(\tilde x , \tilde  \tau)  \Big  \rangle_{\rm con} 
 & =   \frac{\tilde{A}^2\, (-1)^{x+y}} {2}
 \left \langle e^{ -i\,\sqrt{ 2\pi }\,\theta(x,\tau)} \, 
 e^{ i\,\sqrt{ 2\pi }\,\theta(y,0)}  \, \cos\left(  \sqrt{ 8 \,\pi}\,\phi (\tilde{x},\tilde{\tau}) \right) \right  \rangle \nn
& \qquad -G^0(x,y,\tau) \left\langle 
\cos\left(  \sqrt{ 8 \,\pi}\,\phi (\tilde{x},\tilde{\tau}) \right) \right  \rangle,
 \end{align}
 which we will evaluate in real time.

For this calculation, let us first simplify the generic operator
 \begin{align}
   e^{ i\,\alpha\,\sqrt{4\pi}\,\theta(x,t)} \, e^{ -i\,\alpha\,\sqrt{4\pi}\,\theta(y,0)} 
  \, e ^{i\, \beta\,\sqrt{4\pi}\,\phi (\tilde{x},\tilde{t}) } 
 %%%%%%%%%%%%%%%%%%%%%%%%%%%%%%%%%%%%%%%%%%%%%%%%%%%%%%%%%%%%%%%%%%%%%%%%%%%%% 
& \simeq
 \left[ \frac
 {  4\,   x \,y    } 
 { \left\lbrace   \left( x+y \right)^ 2-v^2 \,t^2 \right \rbrace
 	%%%%%
\left \lbrace \left( x-y \right)^ 2-v^2 \,t^2 \right \rbrace} \right ]^{\alpha^2} 
 %%%
  \frac
 { e^{ - \alpha\,\beta\,C_3 (\tilde{x},x,\tilde{t}-t)}   e^{  \alpha\,\beta\,C_3 (\tilde{x},y,\tilde{t})} }
 { (2\, \tilde{x})^{ \beta^2} } 
\nn &  \qquad 
 \times :e^{ i\,\alpha\,\sqrt{4\pi} \,\theta(x,t)} \, 
 e^{ -i\,\alpha\,\sqrt{4\pi}\, \theta(y,0)}  \, 
e ^{i\, \beta\,\sqrt{4\pi}\, \phi (x_1,t_1) }:\,,
 \end{align}
 where
 \begin{align}
 C_3(\tilde{x},y,\tilde{t}) &=
   \sum_{\ell,\ell^\prime}\frac{e^{i\, \omega_\ell \tilde{t}}}
   {\sqrt{\ell \, \ell^\prime}}
  \, [B_\ell(y),A^\dagger_{\ell^\prime}(\tilde{x})]
 = 4\, i\sum_\ell \frac{e^{i \,\omega_\ell\, \tilde{t}}}
 { \ell} \cos \left(\frac{\ell\,\pi \,y}{L}\right)
 \sin \left(\frac{\ell\,\pi \,\tilde{x}}{L}\right) \nn
 %%%%%%%%%%%%%5
 & \simeq \,
 \ln \left[
 \frac{ \left \lbrace  \left( \tilde{x} +y \right) - v  \,\tilde{t}  \right \rbrace 
 	\left \lbrace  \left( \tilde{x}-y \right)-v  \,\tilde{t} \right \rbrace 
 }
 %%%%%
 {  \left \lbrace  \left( \tilde{x}-y \right)+v  \,\tilde{t} \right \rbrace 
 	\left \lbrace  \left( \tilde{x} + y \right)+v  \,\tilde{t} \right \rbrace 
 }
 \right ] .
 \end{align}

Using the above, we find that
 \begin{align}
  e^{ i\,\alpha\,\theta(x,t)} \, e^{ -i\,\alpha\,\theta(y,0)}  \, 
  e ^{i\, \beta\,\phi (\tilde{x},\tilde{t}) } 
&  
\simeq \,\,
:  e^{ i\,\alpha\,\theta(x,t)} \, e^{ -i\,\alpha\,\theta(y,0)}  
\, e ^{i\, \beta\,\phi (\tilde{x},\tilde{t}) }:
\left[ \frac
 {  4\,   x \,y    } 
 { \left \lbrace  \left( x+y \right)^ 2-v^2 \,t^2\right \rbrace
 	%%%%%
 \left \lbrace  \left( x-y \right)^ 2-v^2 \,t^2\right \rbrace} \right ]^{\alpha^2} 
 %%%
 \frac{ 1}{ (2\, \tilde{x})^{ \beta^2} } 
 \nn
 %%%%
 \nn & \quad \times 
 \left[
 \frac{ \left \lbrace  \left( \tilde{x} -y \right) - v  \,\tilde{t}  \right \rbrace 
 	\left \lbrace  \left( \tilde{x}+y \right)-v  \,\tilde{t} \right \rbrace 
 }
 %%%%%
 {  \left \lbrace  \left( \tilde{x}-y \right)+v  \,\tilde{t} \right \rbrace 
 	\left \lbrace  \left( \tilde{x} + y \right)+v  \,\tilde{t} \right \rbrace 
 }
 %%%%
 \frac
 {  \left \lbrace  \left( \tilde{x}+x \right)+v  \left ( \tilde{t}-t \right ) \right \rbrace 
 	\left \lbrace  \left( \tilde{x} - x \right)+v  \left ( \tilde{t}-t \right ) \right \rbrace 
 }
 { \left \lbrace  \left( \tilde{x} +x \right) - v   \left ( \tilde{t}-t \right )  \right \rbrace 
 	\left \lbrace  \left( \tilde{x}-x \right)-v  \left ( \tilde{t}-t \right ) \right \rbrace 
 }
 \right ] ^{\alpha\,\beta}\,.
 \end{align}
 Finally, setting $\alpha=-\frac{1}{\sqrt{2}}$ and $\beta=\pm \sqrt{2}$, and continuing to imaginary time $t\to -i\, \tau$, we obtain
 \begin{align}
&  \frac{\pi}{v}\Big \langle T \,
  S^+(x,\tau)\,S^-(y,0)\,
  \mathcal{O}_2(\tilde x , \tilde  \tau)  \Big  \rangle_{\rm con} 
 = \frac{\tilde{A}^2 (-1)^{x+y}}{16 \,\tilde{x}^2} \sqrt{\frac
   {  4\,   x \,y    } 
   { 
   	\left (   \left( x+y \right)^ 2+v^2 \,\tau^2
   	\right )
   	%%%%%
   	\left (  \left( x-y \right)^ 2+v^2 \,\tau^2
   	\right )}  }
%%%
    \nn 
& \hspace{1.5 cm} \times  \left[
   \frac
   {  \left \lbrace  \left( \tilde{x}-y \right)-  i\,v  \,\tilde{\tau} \right \rbrace 
   	\left \lbrace  \left( \tilde{x} + y \right)-  i\,v  \,\tilde{\tau} \right \rbrace  }
   { \left \lbrace  \left( \tilde{x} -y \right) +i\, v  \,\tilde{\tau}  \right \rbrace 
   	\left \lbrace  \left( \tilde{x}+y \right)+i\,v  \,\tilde{\tau} \right \rbrace 
   }
   \frac
   { \left \lbrace  \left( \tilde{x} +x \right) +i\, v   \left ( \tilde{\tau}- \tau \right )  \right \rbrace 
   	\left \lbrace  \left( \tilde{x}-x \right)  +  i\,v  \left ( \tilde{\tau}- \tau \right ) \right \rbrace 
   }
   {  \left \lbrace  \left( \tilde{x}+x \right)  - i\,v  \left ( \tilde{\tau}- \tau \right ) \right \rbrace 
   	\left \lbrace  \left( \tilde{x} - x \right)-i\,v  \left ( \tilde{\tau}- \tau \right ) \right \rbrace 
   } -1\right]  + \mathrm{H.c.}  
  \end{align}
 It is important to note that this result does not have a bulk limit, and all contributions come only from the boundary limit terms. Furthermore, we see that all the complex numbers in the second line can be expressed by pure phase factors. Therefore, there are no singularities generating logarithmic corrections.

 \end{widetext}

%%%%%%%%%%%%%%%%%%%%%%%%%%%%%%%%%%%%%%%%%%%%%%%%%%

\end{appendix}

%\bibliography{biblio}

\end{document}